\documentclass[pra,twocolumn,superscriptaddress,longbibliography]{revtex4-1}

\usepackage{color}
\usepackage{amssymb,amsmath,amsfonts,bm}
\usepackage{epsfig,graphicx}
\usepackage{epstopdf}
\usepackage{times}
\usepackage{bm}

\begin{document}
\title{Control of the magnon-photon level attraction in a planar cavity}

\author{Y. Yang}
\affiliation{Department of Physics and Astronomy, University of Manitoba, Winnipeg R3T 2N2, Canada}
\affiliation{State Key Laboratory of Infrared Physics, Chinese Academy of Science, Shanghai 200083, China}
\author{J. W. Rao}
\email{jinweir@myumanitoba.ca}
\affiliation{Department of Physics and Astronomy, University of Manitoba, Winnipeg R3T 2N2, Canada}
\author{Y. S. Gui}
\affiliation{Department of Physics and Astronomy, University of Manitoba, Winnipeg R3T 2N2, Canada}
\author{B. M. Yao}
\affiliation{State Key Laboratory of Infrared Physics, Chinese Academy of Science, Shanghai 200083, China}
\author{W. Lu}
\affiliation{State Key Laboratory of Infrared Physics, Chinese Academy of Science, Shanghai 200083, China}
\author{C.-M. Hu}
\email{hu@physics.umanitoba.ca}
\affiliation{Department of Physics and Astronomy, University of Manitoba, Winnipeg R3T 2N2, Canada}

\begin{abstract}
A resistive coupling circuit is used to model the recently discovered dissipative coupling in a hybridized cavity photon-magnon system.  With this model as a basis we have designed a planar cavity in which a controllable transition between level attraction and level repulsion can be achieved. This behaviour can be quantitatively understood using an LCR circuit model with a complex coupling strength. Our work therefore develops and verifies a circuit method to model level repulsion and level attraction and confirms the universality of dissipative coupling in the cavity photon-magnon system. The realization of both coherent and dissipative couplings in a planar cavity may provide new avenues for the design and adaptation of dissipatively coupled systems for practical applications in information processing.
\end{abstract}

\maketitle
\section{Introduction}
Strong light-matter interactions are an interesting and important subject of condensed matter physics, enabling new insight into material characteristics and device design\cite{khitrova1999nonlinear,raimond2001manipulating,xiang2013hybrid,aspelmeyer2014cavity,hu2015dawn,harder2018solid,sharma2018optical}.
Of key importance is the phenomena of Rabi splitting -- the removal of an energy degeneracy due to hybridization, which offers new possibilities for coherent manipulation. Numerically, the vacuum Rabi splitting is twice the product of the transition dipole moment and the vacuum field arising from the root-mean-square of the vacuum fluctuations\cite{khitrova2006vacuum}. To date, the coherent interaction between confined electromagnetic fields and a qubit\cite{wallraff2004strong,majer2007coupling,Sillanpaa}, quantum dot\cite{hennessy2007quantum}, mechanical oscillator\cite{groblacher2009observation,Song}, and magnon\cite{soykal2010strong,huebl2013high,tabuchi2014hybridizing} has been demonstrated. In particular, due to the low room temperature damping rate of microwave photon and magnon and the maturation of microwave technology, the cavity-magnon-polariton (CMP)\cite{bai2015spin} has been brought to the forefront, providing an interesting platform for the merging of quantum electrodynamics and magnetism. Recent progress has demonstrated ultra-strong coupling\cite{goryachev2014high,zhang2014strongly,kostylev2016superstrong}, gradient memory architectures\cite{zhang2015magnon}, the control and readout of qubit states\cite{tabuchi2015coherent}, spin pumping manipulation\cite{bai2015spin,maier2016spin}, and cooperative polariton dynamics\cite{yao2017cooperative}.

At microwave frequencies such hybrid circuits involving charges, spins, and solid-state devices can be fabricated on a chip and integrated with well established microwave technologies, which is crucial for the future development of information processing\cite{xiang2013hybrid}. The engineer-ability of coherent coupling has been made feasible by the development of phenomenologically equivalent LCR modes for such circuits. Depending on the dominant electric or magnetic nature of the confined electromagnetic field, strong light-matter interactions have been modelled by introducing mutual capacitance for a phase qubit\cite{majer2007coupling,pirkkalainen2013hybrid}, quantum dot\cite{delbecq2011coupling} and optomechanical device\cite{singh2014optomechanical} or mutual inductance for a flux qubit\cite{chiorescu2003coherent,abdumalikov2008vacuum} and CMP system\cite{kaur2016chip}. This approach successfully reproduces the key physical phenomena associated with coherently coupled systems, while also enabling on-chip integration.

However, the coupling between light and matter is not strictly limited to coherent interactions. Very recently dissipative coupling induced level attraction has been experimentally discovered, where hybridized modes coalesce rather than repel, due to a Lenz-like effect in a Fabry-Perot cavity\cite{harder2018level}. Such behaviour cannot be described by mutual capacitive or inductive mechanisms. Therefore device integration requires a more general equivalent LCR model. While the physical meaning is very different, from a mathematical point of view level attraction and level repulsion are equivalent to each other through frequency and damping exchange in the plane of complex eigenvalues\cite{okolowicz2003dynamics,bernier2018level,Grigoryan}. For level repulsion the eigenfrequencies, corresponding to the real eigenspectrum, are repelled while the damping of the hybridized modes, determined by the imaginary eigenspectrum, are attracted. The opposite is true for level attraction. This relationship hints at a more comprehensive LCR circuit model which includes both repulsion and attraction -- the imaginary coupling strength required for level attraction should be produced by a mutual resistance that accounts for the dissipative coupling, while the real coupling strength which leads to level repulsion will arise from a mutual capacitance or inductance \cite{pippard2007physics,Tobar}.

Typically the mutual resistive coupling is concealed behind the dominant capacitive and inductive mechanisms. However in this paper we couple YIG ($\mathrm{Y_3Fe_5O_{12}}$) to a specially designed planar microstrip cross junction which enables both level attraction and repulsion. By tuning the YIG position we can manipulate the local rf field distribution and transition between level repulsion, with inductance-dominated coupling, and level attraction, with resistance-dominated coupling. Level attraction in such an on-chip device may provide new avenues for the integrability and practical design of information processing.

\section{Experiment}
A picture of the microstrip cross junction cavity is depicted in Fig. \ref{fig:1} (a). Details of the design and characterization of this cavity is given in Appendix A. During our experiment the $x$-$y$ plane of this cavity is fixed inside an electromagnet which provides an external magnetic field ${H}$ along the ${z}$ direction. To observe photon-magnon coupling, a 1-mm diameter YIG sphere, chosen for its high spin density, low losses and therefore large photon-magnon coupling\cite{serga2010yig}, is mounted on an $x$-$y$-$z$ stage at a fixed height of D = 0.7 mm from the cavity in the ${z}$ direction. This setup allows us to continuously tune the YIG position in the $x$-$y$ plane, and hence to change the local field and the coupling effect. In our experiment the YIG sphere (black circle) can be moved within the 10 mm $\times$ 10 mm range of the dashed blue box shown in Fig. \ref{fig:1} (a). The cavity transmission spectra ${S_{21}}$ (green symbols) is displayed in Fig. \ref{fig:1} (c), where three resonant modes, with frequencies of ${\omega_c}/2\pi$ = 3.22, 6.253 and 9.39 GHz, are labelled as Mode 1, Mode 2 and Mode 3. A detailed discussion about the cavity resonance can be found in Appendix A. Black and red curves correspond to theoretical calculations using Eq. \eqref{SM2b} and Computer Simulation Technology (CST) simulations, respectively.

\begin{figure}
\begin{center}\
\epsfig{file=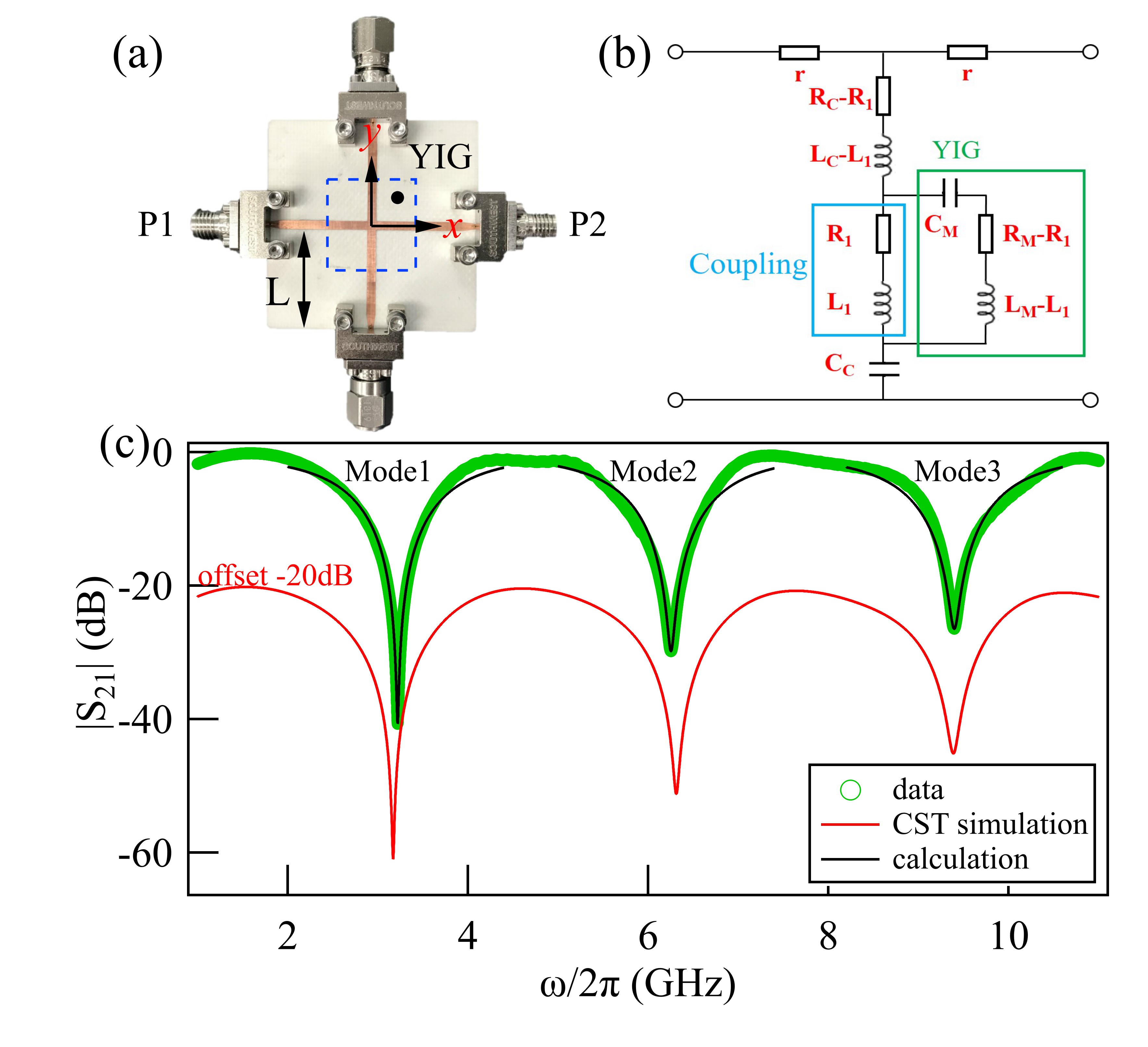,width=8.5 cm}
\caption{\label{fig:1}(a) Picture of the measurement setup with a 1-mm diameter YIG sphere placed a height D = 0.7 mm above the planar microstrip cross junction cavity. The two short-terminated vertical arms and two horizontal arms each have a length of L = 20 mm. The planar cavity is placed in the ${x-y}$ plane while an external magnetic field ${H}$ is applied in the ${z}$ direction (perpendicular to the planar cavity).  The range of YIG locations is indicated by the 10 mm $\times$ 10 mm dashed blue box. (b) Equivalent circuit of the coupled system. Circuit elements used to model the YIG sphere are highlighted by the green box while the coupling term is emphasized by a blue box. (c) Cavity spectra of experimental data (green), theory (black) and CST simulation (red, -20 dB offset), with three resonant modes labelled Mode 1, Mode 2 and Mode 3.}
\end{center}
\end{figure}

Essentially, the coupling mechanism of coupling effects in the coupled cavity photon magnon system can be defined by three fundamental electrodynamic principles: Amp\`ere's Law, Faraday's Law and Lenz's Law \cite{harder2018level}. Specifically, the inductive current of the cavity induces a magnetic field (Amp\`ere's Law), which applies a driving torque to the spin in the magnon system (the YIG sphere). Because of the spin precession, the magnetic flux of the cavity is altered. As a consequence, an induced current is generated in the cavity circuit (Faraday's Law), which affects the dynamic properties of the cavity mode. Usually these two principles dominate, leading to a coherent coupling in the cavity-photon magnon system. However, in some special cases such as in the cross cavity, another electrodynamic principle, Lenz's Law, must also be considered.  In such cases an additional magnetic field is generated by the induced current of the spin procession, which applies a drag torque and tends to impede the spin procession (Lenz's Law). Therefore, we conclude that the coupling is coherent if the driving torque due to Amp\`ere's Law is dominant over the drag torque from Lenz's Law.  Otherwise, the coupling effect becomes dissipative.

In order to demonstrate the significance of these principles for concrete applications, in this paper we propose a phenomenological LCR model for both coherent and dissipative coupling in different physical systems. The equivalent LCR circuit, specific to the coupled cavity-photon magnon system reported in this paper, is shown in Fig. \ref{fig:1} (b). To couple with the cavity circuit, the YIG sphere acts as a resonant circuit, with self inductance, capacitance and resistance connected in series and labelled as ${L_m, C_m, R_m}$. To describe the coupling we consider both direct and indirect interactions between the cavity electromagnetic fields and the magnetic material.  First, the electromagnetic field will be directly influenced by the permittivity and permeability, which in general can be modelled by a mutual capacitance and inductance respectively \cite{barry1986broad,Tobar,hong2004microstrip,kaur2016chip}. Due to the magnetic properties of YIG we can describe the coherent coupling effects due to the resonant permeability through a mutual inductance. In other words, the rf cavity current will produce a magnetic field, which drives magnetization precession and induces a voltage in the cavity circuit. This coupling between the cavity current and the YIG is characterized by the mutual inductance, which is the ratio between the induced YIG voltage and the time derivative of the varying cavity current\cite{grover2004inductance,zangwill2013modern}. However, there is also an indirect interaction whereby the induced electric field of the YIG sphere will produce an additional cavity current due to the finite cavity conductivity. This coupling is related to damping and energy dissipation. It may be modelled by a mutual resistance\cite{Tobar,hurley1995calculation}, and could be realized by the 90 degree phase lag in the coupling term when compared with the coherent coupling. Therefore we choose a combination of mutual inductance ${L_1}$ and mutual resistance ${R_1}$ to describe both the magnetic inductive coupling of level repulsion and the resistive coupling of level attraction\cite{pippard2007physics}. The equivalent circuit of the coupled system is shown in Fig. \ref{fig:1} (b).

\begin{figure*}
\begin{center}\
\epsfig{file=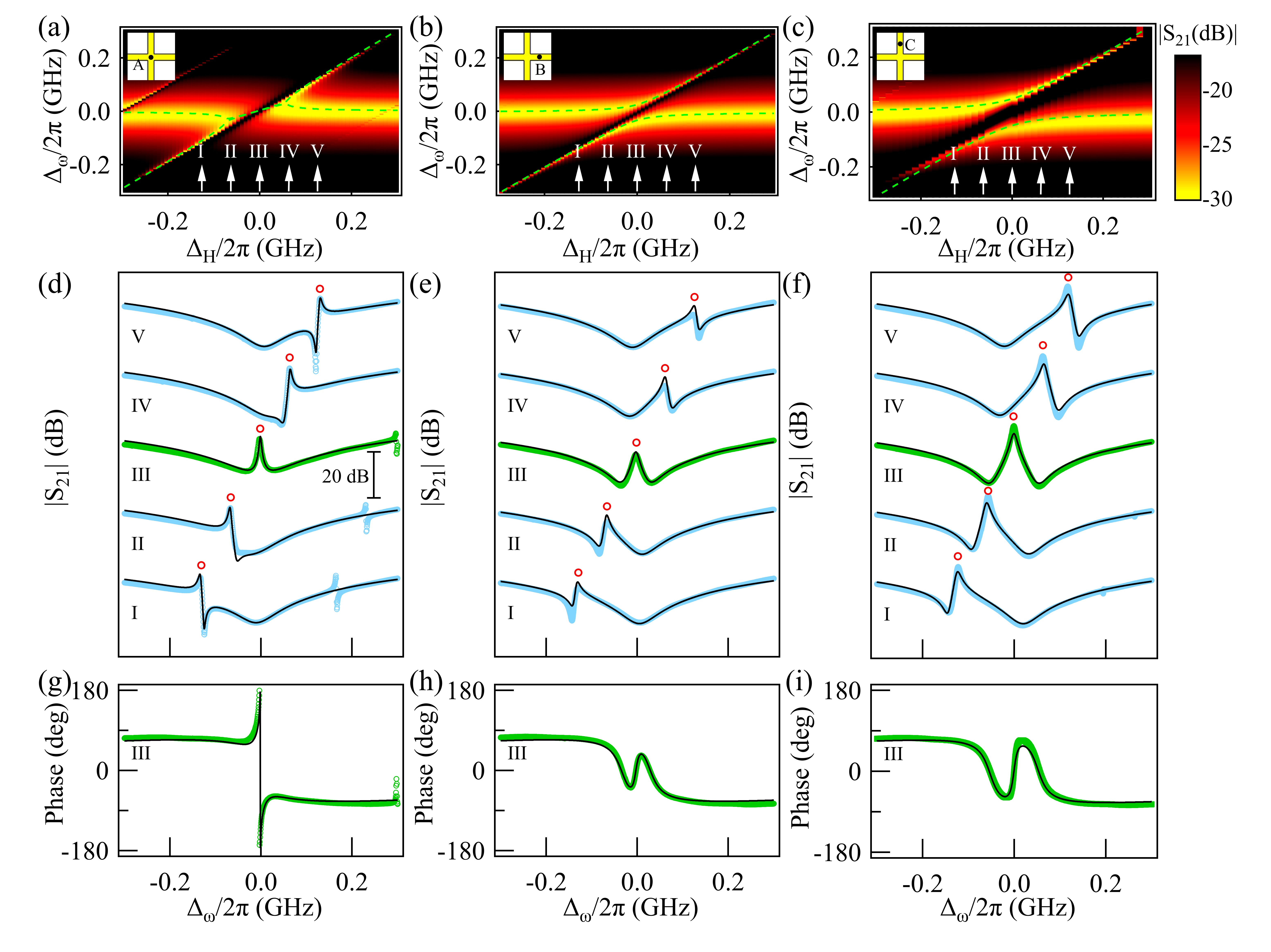,height=10.6 cm}
\caption{\label{fig:2} (a) S$_{21}$ amplitude mapping, (d) S$_{21}$ amplitude spectra and (g) S$_{21}$ phase spectra of level attraction when YIG is mounted in position A. The same for (b), (e) and (h) in position B and (c), (f) and (i) in position C. Amplitude peaks are labelled as red circles in the amplitude spectra. The green dash lines in the amplitude mappings and black curves in the amplitude and phase spectra are theoretical calculations.}
\end{center}
\end{figure*}

Using the LCR circuit model the transmission spectra of the magnon-photon coupled system can be calculated as,
\begin{equation}
\text{S}_{21}\propto 1-\frac{i\gamma_{ce}}{\omega-\omega_c+i(\gamma_{ce}+\gamma_{ci})+\frac{G^2}{\omega-\omega_m+i\gamma_m}}
\label{one}
\end{equation}
where ${\omega_c = {1}/{\sqrt{L_cC_c}}}$ and ${\omega_m = {1}/{\sqrt{L_mC_m}}}$ are the resonance frequencies of the cavity and magnon, respectively. ${\gamma_m = {R_m}/2L_m}$ is the YIG damping while ${\gamma_{ce}}$ and ${\gamma_{ci}}$ are the extrinsic and intrinsic cavity damping, see Appendix A for detailed discussion of cavity transmission.  The complex coupling strength ${G = {(L_1\omega+iR_1)}/{\sqrt{4L_cL_m}}}$ is related to the mutual inductance and resistance. For level repulsion ${L_1}$ is dominant, the coupling strength is real and the hybridized modes are repelled. For level attraction ${L_1}$ diminishes and results in an imaginary coupling strength due to ${R_1}$. In this case the modes are attracted by the coupling.  In order to be consistent with the notation of Ref. \citenum{harder2018level} we choose an absolute value of the coupling strength ${|G|=|g e^\frac{i\Phi}{2}|}$. Therefore for pure level repulsion ${\Phi = 0}$ and the absolute coupling strength is ${|g| = {L_1\omega}/{\sqrt{4L_cL_m}}}$, while for pure level attraction ${\Phi = \pi}$ and ${|g| = {R_1}/{\sqrt{4L_cL_m}}}$. Experimentally, level repulsion exhibits as two hybridized modes of the system repel each other in the frequency domain, when the coherent coupling is dominant. By contrast, during level attraction the two hybridized modes coalesce with each other and occur when dissipative coupling becomes dominant.

An amplitude mapping of the microwave transmission spectra S$_{21}$, measured using a vector network analyzer with the YIG in the centre of the cross junction at position A, is shown in Fig. \ref{fig:2} (a) as a function of the frequency and field detunings, ${\Delta_{\omega} = \omega-\omega_c}$ and ${\Delta_{H} = \omega_r(H)-\omega_c}$. Here we have used Mode 2 with ${\omega_c/2\pi = 6.253}$ GHz. The uncoupled magnon mode with damping ${\gamma_m/2\pi = 0.004}$ GHz follows the Kittel dispersion ${\omega_r(H)=\gamma(H+H_A)}$, where ${\gamma=2\pi\times27.4}$ $\mu_0$GHz/T is the gyromagnetic ratio and ${\mu_0H_A = 2.26}$ mT is the magneto-crystalline anisotropy field.

At position A, a high symmetric point of the structure, where both the rf e-field and h-field are at a minimum no matter which port is input, see Fig. \ref{fig:SM2} (c) and (f), we observe level attraction by tuning the FMR frequency around ${\omega_c}$. The hybridized frequencies bend towards each other and meet at two exceptional points\cite{bernier2018level, harder2018level}. In the region between these two points the modes coalesce and the absolute coupling strength is ${|g/2\pi|}$ = 33 MHz. In order to change the coupling feature we moved the YIG in the $x$-$y$ plane to position B and C. The field distribution at position B is shown in Fig. \ref{fig:SM2} (c) and (f). Although no field is input at Port 2 while measuring S$_{21}$, the vacuum field couples to the magnons and leads to magnon-photon Rabi oscillations\cite{agarwal_2012}. As shown in Fig. \ref{fig:2} (b) and (c), level repulsion is observed at position B (${x = 1.82}$ mm) in the right arm and at position C (${y = 1.82}$ mm) in the upper arm. In these cases the two hybridized modes are repelled by each other and open a Rabi-like gap, with a coupling strength of 32.5 and 54 MHz respectively, determined from the splitting.

In Fig. \ref{fig:2} (d) when the field detuning is set to ${\Delta_{H} = 0}$, the transmission spectra amplitude of level attraction is plotted as a function of ${\Delta_{\omega}}$ using green circles.  A resonance peak appears at ${\Delta_{\omega} = 0}$.  However it looks quite similar to the zero-detuning spectrum of level repulsion in Fig. \ref{fig:2} (e) and (f).
Fortunately the amplitude peaks in our system, characterized by a suppression of $|$S$_{21}|$ amplitude originating from the destructive interference between the magnon response and driving force\cite{harder2016spin}, can be used to distinguish the two forms of coupling.  For the known case of level repulsion, shown in Fig. \ref{fig:2} (e) and (f), the amplitude peak always appears between the coupled modes. However in the case of level attraction the peak appears outside of the two hybridized modes.

Another robust method to distinguish level attraction and repulsion is to examine the transmission phase at ${\Delta_{H} = 0}$. In level attraction a single ${2\pi}$-phase jump at ${\Delta_{\omega}=0}$ is observed in Fig. \ref{fig:2} (g), corresponding to an amplitude peak between the two attracted modes. In level repulsion the amplitude dip at each hybridized mode corresponds to two ${\pi}$ phase delays, while the peak in between is observed as an opposite ${\pi}$-phase shift at ${\Delta_{\omega}=0}$ \cite{sames2014antiresonance,harder2016spin}, see Fig. \ref{fig:2} (h) and (i). Both techniques are in agreement and confirm the presence of level attraction in the transmission spectra of Fig. \ref{fig:2} (d).

\begin{figure}
\begin{center}\
\epsfig{file=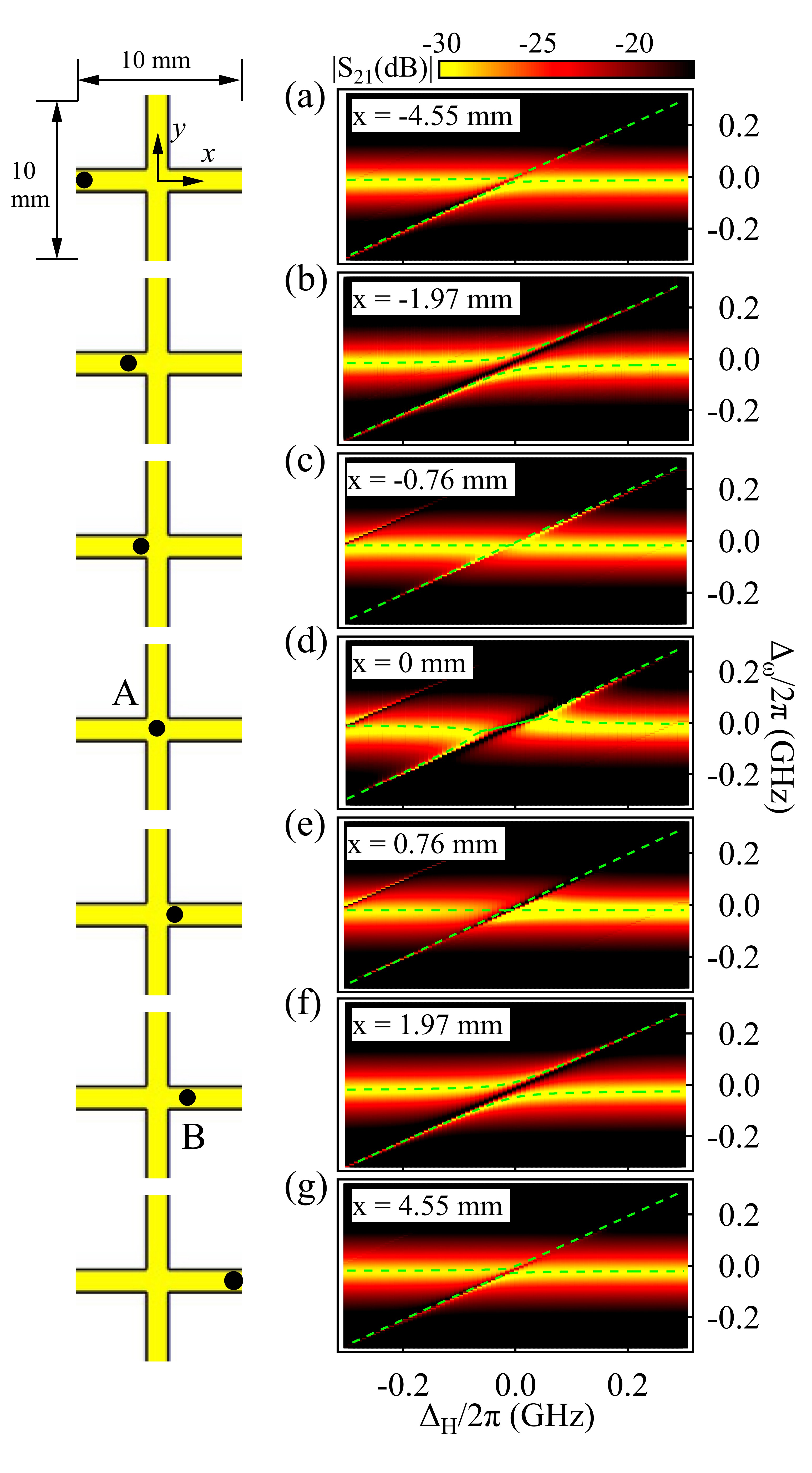,width=8 cm}
\caption{\label{fig:3} In the left panel, positions of YIG are given within a 10 mm ${\times}$ 10 mm area. Amplitude mappings in the right panel show a systemtic evolution from level repulsion to level attraction and back to level repulsion. The green dash lines in amplitude mapping are theoretical calculations.}
\end{center}
\end{figure}

To examine the transition between level attraction and repulsion we moved the YIG position continuously along the $x$ axis, between $|x| <$ 5 mm.  Selected YIG positions and the corresponding mappings are shown in Fig. \ref{fig:3}, demonstrating the systematic evolution between level repulsion and level attraction about the crossing point (${x}$ = 0 mm, ${y}$ = 0 mm).
When the ${|x|}$ position is decreased from ${|x| = 4.55}$ mm to ${|x| = 0.76}$ mm the inductive coupling is dominant and the mapping shows level repulsion.  The Rabi-like gap between the two hybridized modes increases at first, reaching a maximum at ${|x| = 1.97}$ mm, after which the gap gradually closes, reaching a minimum at ${|x| = 0.76}$ mm where the two modes appear to cross. After ${|x| = 0.76}$ mm the system enters a level attraction region, dominated by resistive coupling, and the coupling strength again increases.  At ${x = 0}$ mm the strongest level attraction is observed.

\begin{figure}
\begin{center}\
\epsfig{file=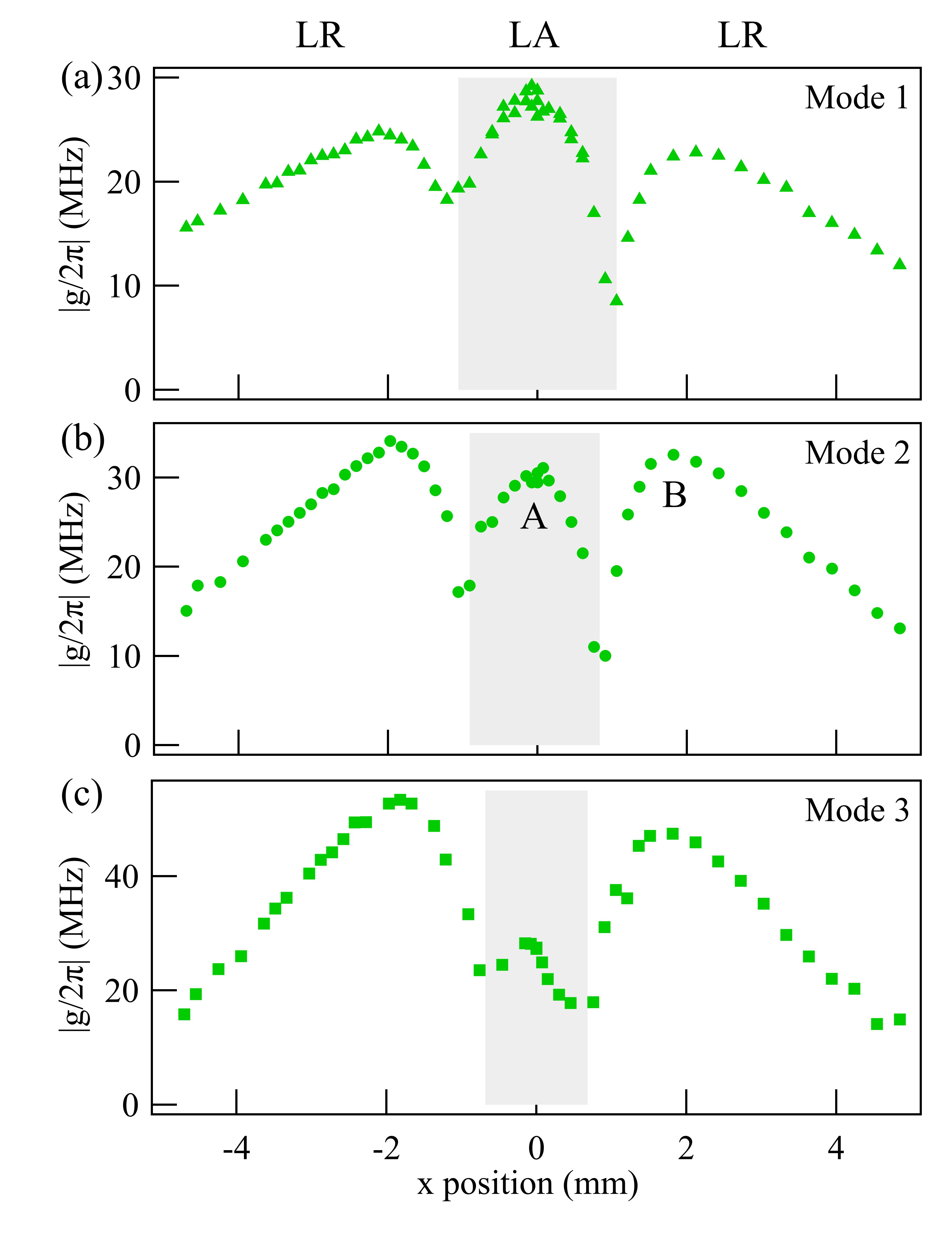,width=8.5 cm}
\caption{\label{fig:4} Evolution of the absolute coupling strength, $|g|$, for (a) Mode 1, (b) Mode 2, and (c) Mode 3 when YIG is moved from ${x}$ = -5 mm to 5 mm with ${y}$ = 0 mm.  The shaded region indicates level attraction. LR and LA are abbreviations of level repulsion and level attraction.}
\end{center}
\end{figure}

The coupling strength depending on YIG sphere's $x$ position can be determined for all three modes by using Eq. (\ref{one}) to fit S$_{21}$.  The results are summarized in Fig. \ref{fig:4}. In the shaded region the system is dominated by resistive coupling, where level attraction is observed.  This is in contrast to the region of level repulsion dominated by inductive coupling.  At ${|x|}$ = 5 mm a small inductance leads to weak level repulsion.  As ${|x|}$ decreases the Rabi-like gap gradually opens and the coupling strength increases to its maximum value, meaning that the inductance must increase and the interaction proceeds via the magnetic field. After that the mutual inductance begins to decrease and then the mutual resistance ${R_1}$ emerges and grows in near level crossing condition. When the overall coupling strength reaches a minimum, a level crossing appears and marks the transition.  As $|x|$ is decreased further beyond the crossing condition towards zero, level attraction is observed, with maximal effect at ${|x|}$ = 0 due to the indirect interaction between the photon and magnon modes. Consistent with previous study using a special Fabry-Perot-like resonator\cite{harder2018level}, it is clearly seen that two competing magnon-photon coupling effects coexist at general experimental conditions in our planar cavity.

\section{Conclusions}
In this work we have developed an LCR circuit model to describe both level repulsion and its transition into level attraction, and have experimentally demonstrated the existence of mutual resistive coupling induced level attraction in a planar cavity.  The realization of resistive coupling provides a new avenue for the development of circuit designs which implement the phenomenon of level attraction. By realizing such an on-chip device, future coupling modules may be more easily integrated into a lumped element system.

\textit{Note added.}
After the paper was written, we found in the last week a preprint also studying the level attraction effect in planar cavity, but with a different cavity design, see arXiv:1901.01729, 2019.

\acknowledgements
This work was funded by NSERC and the China Scholarship Council. We would like to thank M. Harder, I. Proskurin, R. L. Stamps and T. J. Silva for helpful discussions and suggestions.

\appendix
\renewcommand\thefigure{\thesection.\arabic{figure}}
\renewcommand{\thetable}{\thesection.\arabic{table}}
\renewcommand{\theequation}{\thesection.\arabic{equation}}
\section{Design and characterization of cavity}
\setcounter{figure}{0}

\begin{figure}[htb]
\begin{center}\
\epsfig{file=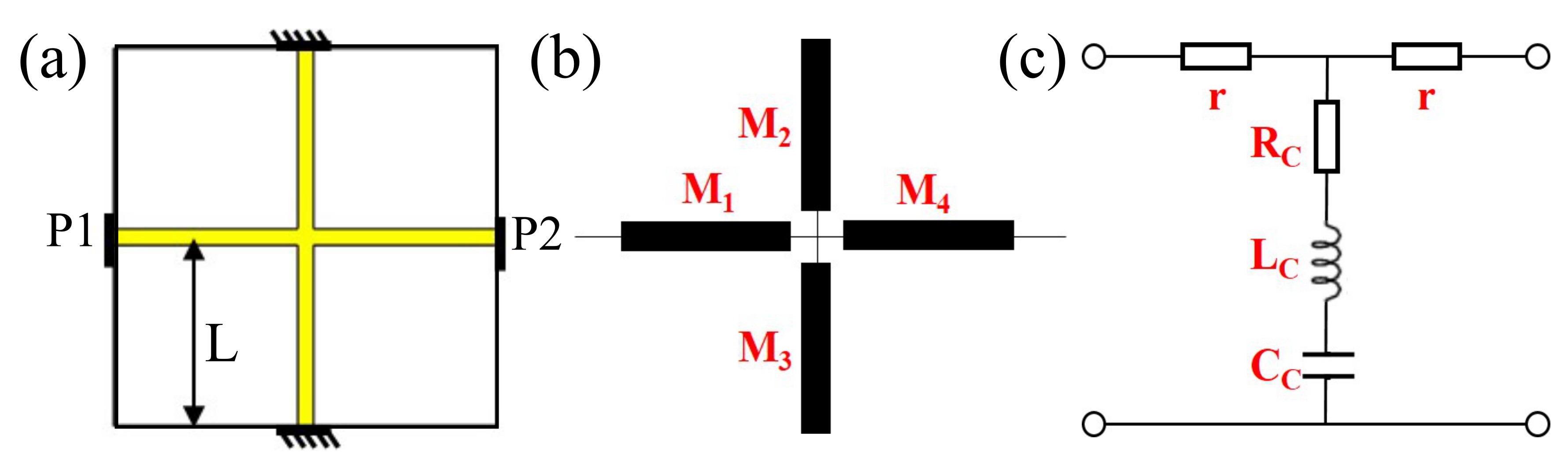,width=8.5 cm}
\caption{\label{fig:SM1} (a) Design of a Michelson-type microwave interferometer with two short-terminated vertical arms and two horizontal arms. (b) The ideal topology of the interferometer consists of two series transmission lines and two short-ended shunt stubs with the same impedance and electrical length.  The individual ABCD matrices are given by ${M_1, M_2, M_3, M_4}$.  (c) Equivalent circuit of the interferometer. }
\end{center}
\end{figure}

Inspired by the well known performance of interferometric techniques, which can operate over a large frequency range and have excellent signal-to-noise ratio (SNR) in magnetic resonance experiments\cite{krishna2017dual,edwards2017microwave}, we used microstrip cross junction to fabricate a Michelson-like interferometer. The microstrip cross junction topology is depicted in Fig. \ref{fig:SM1} (a). This cross-shaped microstrip is fabricated using two perpendicular 1.67 mm-wide transmission lines on a 0.813 mm thick RO4003C substrate. The two horizontal arms are connected to a vector network analyzer (VNA) to enable microwave transmission measurements while  two vertical arms are short-terminated, acting as the boundaries of the cavity. Each arm shares the same characteristic impedance ${Z_0= 50~ \Omega}$ and electrical length ${\theta=kL}$. ${L=20}$ mm is the real length of each arm and ${k=i\alpha+\beta}$ is the complex wave number in a lossy material\cite{pozar2009microwave}. Attenuation constant is ${\alpha}$ and phase constant ${\beta=\omega/\upsilon_p}$, $\upsilon_p$ is the phase velocity at medium.

Because the feature dimensions of the cross junction are much smaller than the wavelength of the microwaves employed, the scattering properties of our device can be modelled by the cascade matrices ${M_1, M_2, M_3, M_4}$ shown in Fig. \ref{fig:SM1} (b). To compute the ABCD matrix for the whole cavity, we can simply multiply the matrices of the individual two-port element\cite{edwards2017microwave}:

\begin{subequations}\label{SM1}
\begin{equation}
\begin{split}
\left[
\begin{array}{cc}
A & B \\
C & D \\
\end{array}\right]
  & = M_1M_2M_3M_4  \\
  & =\left[
\begin{array}{cc}
2\cos\left(2\theta\right)+1 & 2jZ_0\sin\left(2\theta\right)\\
-2j\frac{\cos\left(2\theta\right) \cot\theta}{Z_0} & 2\cos\left(2\theta\right)+1 \\
\end{array}\right]
\end{split}
\label{SM1a}
\end{equation}
where
\begin{equation}\label{SM1b}
M_1 = M_4 = \begin{bmatrix}
       \cos\theta & jZ_0\sin\theta  \\
       j\frac{1}{Z_0}\sin\theta & \cos\theta  \\
     \end{bmatrix}
\end{equation}
\begin{equation}\label{SM1c}
M_2 = M_3 = \begin{bmatrix}
       1 & 0  \\
       \frac{1}{jZ_0\tan\theta} & 1  \\
     \end{bmatrix}
\end{equation}
\end{subequations}

From this ABCD matrix, the transmission parameter can be derived as:
\begin{subequations}\label{SM2}
\begin{equation}
%\begin{split}
S_{21}
=\frac{2}{A+B / Z_0+C Z_0+D}
=\frac{1}{2}e^{i2\theta}(1-e^{i2\theta})
\label{SM2a}
%\end{split}
\end{equation}
which results in a resonant dip of the transmission spectra at $\omega=\omega_c$.

For the near resonant condition ${\alpha L \ll 1}$, so ${e^{-2\alpha L}=\Gamma\approx 1}$ is smaller than 1\cite{edwards2017microwave}.  Furthermore ${\beta_c=n\pi/L=\omega_c/\upsilon_p}$ when ${n}$ is an integer, so ${e^{i2\beta_cL}=1}$ and Eq. \eqref{SM2a} can be rewritten as:
\begin{equation}
S_{21}\approx \frac{1}{2}\left[1-\frac{i\gamma_{ce}}{\omega-\omega_c+i(\gamma_{ce}+\gamma_{ci})}\right]
\label{SM2b}
\end{equation}
\end{subequations}
where ${\gamma_{ce}=\Gamma\frac{\upsilon_p}{2L}}$ and ${\gamma_{ci}=(1-\Gamma)\frac{\upsilon_p}{2L}}$ are the extrinsic and intrinsic damping of the cavity.

\begin{figure}
\begin{center}\
\epsfig{file=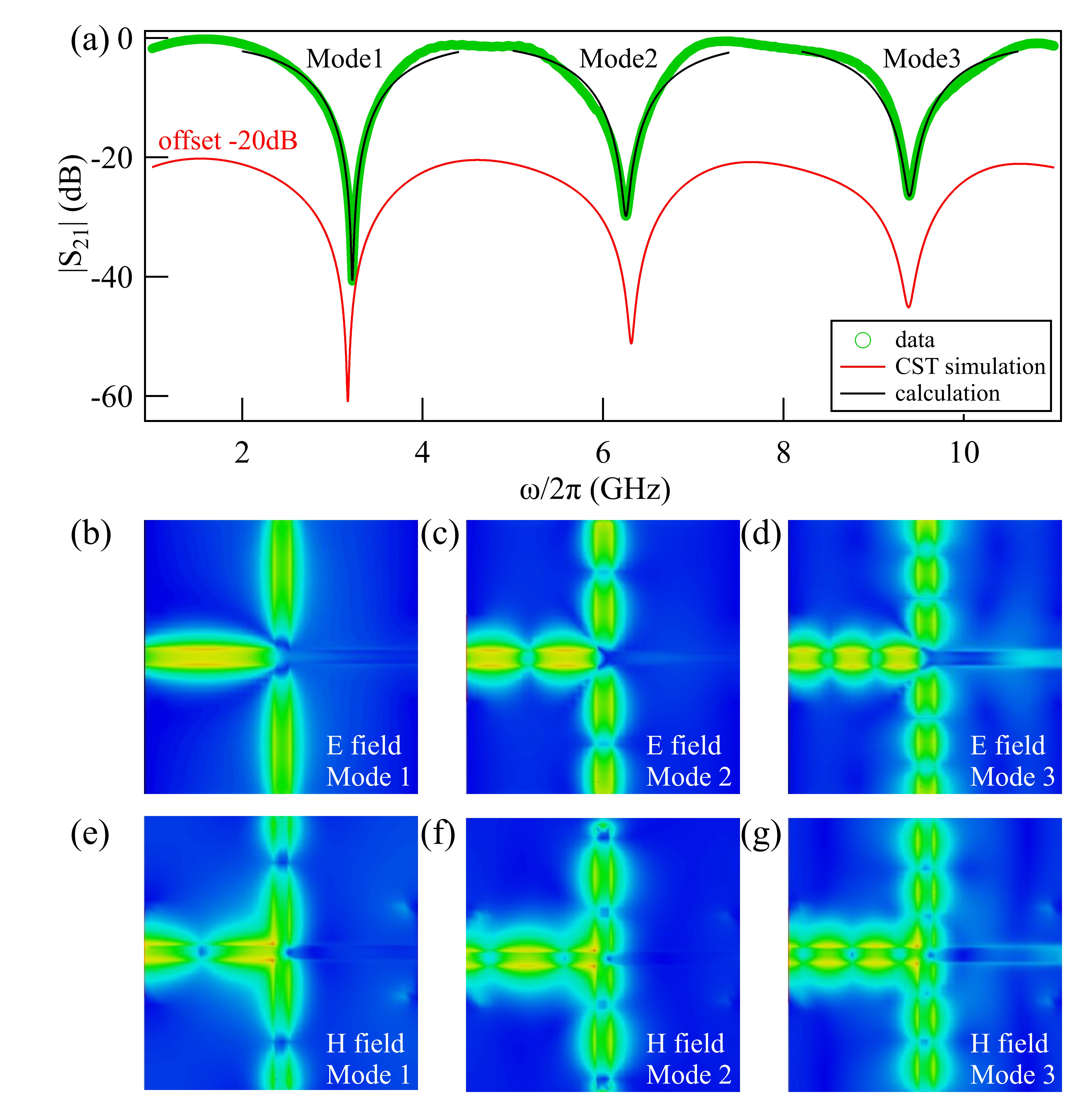,width=8.5 cm}
\caption{\label{fig:SM2}(a) The cavity spectra $S_{21}$ is shown as green symbols, with black and red curves corresponding to the theoretical calculation and CST simulations (-20 dB offset), respectively. The resulting interference fringes of rf electric field of Mode 1, Mode 2 and Mode 3 are shown in (b), (c) and (d) while the rf magnetic field distributions are shown in (e), (f) and (g).}
\end{center}
\end{figure}

\begin{table}
\begin{center}
\caption{Resonance frequency, $\omega_c$, extrinsic damping, $\gamma_{ce}$, and intrinsic damping, $\gamma_{ci}$, of Mode 1, Mode 2 and Mode 3.} \vspace{10pt}
\begingroup
\setlength{\tabcolsep}{10pt} % Default value: 6pt
\renewcommand{\arraystretch}{1.5} % Default value: 1
\begin{tabular}{ c   c  c  c}	
\hline \hline	
  & \textbf{Mode 1} & \textbf{Mode 2} & \textbf{Mode 3} \\ \hline
  $\omega_c/2\pi$ & 3.22 GHz & 6.253 GHz & 9.39 GHz \\
  $\gamma_{ce}/2\pi$ & 0.99 GHz & 0.99 GHz & 0.99 GHz \\
  $\gamma_{ci}/2\pi$ & 0.010 GHz & 0.034 GHz & 0.052 GHz \\
  \hline
\end{tabular}
\endgroup
\end{center}\label{Tab:SM3}
\end{table}

Figure \ref{fig:SM1} (c) shows the equivalent phenomenological LCR circuit model which quantitatively describes the resonant behaviour. The circuit consists of lumped elements of resistance ${r}$, ${R_c}$, inductance ${L_c}$ and capacitance ${C_c}$. The matrix of this circuit is:
\begin{equation}
\left[
\begin{array}{cc}
A & B \\
C & D \\
\end{array}\right]
   =\left[
\begin{array}{cc}
1+\frac{r}{Z_c} & r(2+\frac{r}{Z_c})\\
\frac{1}{Z_c} & 1+\frac{r}{Z_c} \\
\end{array}\right]
\label{SM3}
\end{equation}
This lumped circuit ABCD matrix must correspond to Eq. \eqref{SM1a} which allows us to identify the two small symmetric resistances ${r = jZ_0\tan\theta\ll Z_0}$ which contribute the extrinsic losses and a shunt impedance ${Z_{c} = R_c+j\omega L_c+\frac{1}{j\omega C_c} = j Z_0 \tan\theta/(2 \cos(2\theta))}$ which describes the cavity resonance. Defining ${\omega_c=1/\sqrt{L_cC_c}}$, ${\gamma_{ce}=(Z_0+r)/4L_c}$ and ${\gamma_{ci}=R_c/2L_c}$, the transmission equation derived from the circuit model is the same as Eq. \eqref{SM2b}. We note that
${\gamma_{ci} \ll \gamma_{ce}}$ since ${1-\Gamma\ll\Gamma}$ and ${R_c\ll \frac{1}{2}(Z_0+r)}$.

The performance of this Michelson-type microwave interferometer was first characterized using a VNA measurement; see the green circles in Fig. \ref{fig:SM2} (a).  The three resonant modes are labelled Mode 1, Mode 2 and Mode 3.  Using Eq. \eqref{SM2b} (black curves) the resonant features can be calculated, with the resonant frequency ${\omega_c}$, extrinsic damping ${\gamma_{ce}}$ and intrinsic damping ${\gamma_{ci}}$ determined based on a fit to the experimental data of Fig. \ref{fig:SM2} (a) and summarized in Table \ref{Tab:SM3}.  As expected all modes share the same extrinsic damping while the intrinsic damping increases with resonant frequency, which may be due to interference effects. The resonant features are also reproduced by CST, and are plotted as the red curve with a -20 dB offset. To clearly see the interference pattern, the resulting interference fringes of the rf e-field and h-field, as modelled by CST, are shown in Fig. \ref{fig:SM2} (b) - (g), in which colour variance reflects the absolute electric or magnetic field strength. As a result, a minimum transmission ($S_{21}$) appears.

\end{document}